\documentclass[aps,preprint,amsmath,amssymb]{revtex4}
\usepackage{graphicx}
\usepackage{epstopdf}
\begin{document}
\title{Dimension-six anomalous $tq\gamma$ couplings in $\gamma \gamma$ collision at the LHC}

\author{S. C. \.{I}nan}
\email[]{sceminan@cumhuriyet.edu.tr}
\affiliation{Department of Physics, Cumhuriyet University,
58140, Sivas, Turkey}

\begin{abstract}
We have investigated the flavor changing top quark physics on the dimension-six anomalous $tq\gamma$ ($q=u,c$) couplings through the process $pp\to p\gamma \gamma p\to p t\bar{q}p$ at the LHC by considering different forward detector acceptances. In this paper, we have also examined the effects of top quark decay. The sensitivity bounds on the anomalous couplings and $t\to q\gamma$ branching ratio have been obtained at the $95\%$ confidence level for the effective lagrangian approach. Besides, we have investigated the effect of the anomalous couplings on single top quark spin asymmetry.
\end{abstract}

\maketitle

\section{Introduction}

Top quark mass is at the electroweak symmetry-breaking
scale. It is the heaviest and one of the least known elementary particle in the Standard Model (SM)\cite{ben, cha, wag}. Therefore, the top quark
properties and its interactions provide a possibility for
examining new physics beyond the SM. Moreover, the effects of new physics theories on the top quark interactions are considered to be larger than any other particles \cite{agu}. New physics interactions
would alter top quark production and decay at the colliders. The most widely studied cases are top quark anomalous interactions via
Flavour-Changing Neutral Currents (FCNC). Tree level FCNC
decay $t\rightarrow q\gamma$ ($q=u,c$) is not possible in the SM. This
decay can only make loop contributions and it is highly suppressed due to Glashow-Iliopoulos-Maiani (GIM) mechanism. For this reason,  $t\rightarrow q\gamma$ branching ratio is very small ($\approx{10^{-14}}$) \cite{he,ee1,ee2,8}. In this
instance, conflicts with the SM expectations of this decay would be evidence of new
physics. These kind of decays have been studied in various new physics models beyond the
SM: quark-singlet model \cite{8,9,10}, the two-Higgs doublet model \cite{11,12,13,14,15,16}, the minimal supersymmetric model \cite{17,18,19,20,21,22,23}, supersymmetry \cite{24}, the top-color-assisted technicolor model \cite{25}
or extra dimensional models \cite{26,27}.

Present experimental constraints on the FCNC $tq\gamma$ couplings are the following:
The CDF collaboration limit on the branching ratio at $95\%$ C. L. for the process $t\to q\gamma$ is $BR(t \to u
\gamma)+BR(t \to c\gamma)<\%3.2$ \cite{cdf}.
The ZEUS collaboration provide at $95\%$ C.L. on the
anomalous $tq\gamma$ coupling $\kappa_{tq\gamma}<0.12$ \cite{zeus} with the assumption of $m_{t} = 175$ GeV. The Large Hadron Collider (LHC) can produce top quarks in the order of
millions per year. Therefore, top quark couplings can be probed with very high sensitivity. In particular, both the ATLAS and
CMS collaboration have presented their sensitivity bounds on these rare top quark decays induced by the
anomalous FCNC interactions \cite{he9,he10,he11}. The most stringent experimental bounds recently have been obtained at $95\%$ C.L. by the CMS Collaboration as $BR(t\to u\gamma)=\%0.0161$ and $BR(t\to c\gamma)=\%0.182$ \cite{cms}. The CMS group can be distinguished the $tu\gamma$ and $tc\gamma$ couplings with applying charge ratio method \cite{crt}. Due to the fact that the u-quark parton distribution function is larger than the c-quark, they have found less sensitivity to $tc\gamma$ coupling.

The effects of new physics to FCNC top quark couplings can be
obtained in a model independent way by means of the effective operator formalism.
The theoretical basis of that kind of a method rely on the assumption that the SM of particle physics is the low-energy limit of a more
fundamental theory. Such a procedure is quite general
and independent of the new interactions at the new physics energy scale. According to Buchm\"{u}ller and Wyler \cite{bah}, these effective operators obey the
$SU(2)_L\times U(1)_Y$  gauge symmetries of the SM and can be written in following form,

\begin{eqnarray}
\textit{L}=\textit{L}_{SM}+\frac{1}{\Lambda}\textit{L}^{(5)}+\frac{1}{\Lambda^2}\textit{L}^{(6)}+O\left(\frac{1}{\Lambda^3}\right)
\end{eqnarray}

\noindent where , $\Lambda$ is the energy scale of new physics,  $\textit{L}_{SM}$ is the SM Lagrangian, $\textit{L}^{(5)}$ and $\textit{L}^{(6)}$ are all
of the dimension-five and dimension-six operators. As mentioned before, they are invariant under the gauge symmetries of the SM. The five dimensional terms break the conversation of lepton and baryon numbers. Hence, we do not examine these operators in this paper. The list of $\textit{L}^{(6)}$ terms is quite vast.
In Refs. \cite{fer1, fer2, fer3, fer4, fer5}, the authors have investigated all dimension-six flavor changing effective operators of the $tqg$ (g:gluon) and $tqV$ ($V:\gamma,Z$) FCNC top physics.
In this paper, we examine the dimension-six operators that give rise to flavor changing interactions of the top quark in the electromagnetic interactions. These operators can be written as shown in \cite{fer4, fer5},

\begin{eqnarray}
\textit{O}_{tB}=&&i\frac{\alpha_{it}^B}{\Lambda^2}(\bar{u}_R^i\gamma^\mu D^\nu t_R)B_{\mu\nu}, \nonumber \\
\textit{O}_{tB\phi}=&&\frac{\beta_{it}^B}{\Lambda^2}(\bar{q}_L^i \sigma^{\mu\nu}t_R)\tilde{\phi}B_{\mu\nu}, \nonumber \\
\textit{O}_{tW\phi}=&&\frac{\beta_{it}^W}{\Lambda^2}(\bar{q}_L^i\tau_I \sigma^{\mu\nu}t_R)\tilde{\phi}W_{\mu\nu}^I,
\label{oprs}
\end{eqnarray}

\noindent where $\alpha_{it}^B$, $\beta_{it}^B$ and $\beta_{it}^W$ dimensionless complex coupling constants, $u_R$ and $q_L$
show the right-handed u-quark singlet and left-handed doublet. $B_{\mu\nu}$ and $W_{\mu\nu}^I$ are the $U(1)_Y$ and $SU(2)_L$ field tensors, respectively.  $\phi$ is the SM Higss doublet, $\tau^I$ are the Pauli matrices and $\tilde{\phi}$ charge conjugate
of the Higgs doublet ($\tilde{\phi}=i\tau^2\phi^*$). Obviously, these operators contribute to $t$ quark anomalous FCNC interactions including photon and $Z$ boson when the partial derivative of
$D_\mu$, the Higgs field $\phi$ and it's vacuum expectation value $\nu$ are used in the Eqs. (\ref{oprs}), through the well-known Weinberg rotation. Moreover, the $Z$ boson couple with the Higgs field. Therefore, there are several extra effective operators which
will only contribute to new FCNC interactions of the $Z$ boson \cite{fer1, fer2}. These operators will not be considered
in this paper, since we analyse only $tq\gamma$ anomalous interactions. The FCNC photon and $Z$ boson couplings with $t$-quark
can be isolated defining new coupling constants,

\begin{eqnarray}
\label{cns}
\alpha^\gamma=&&\cos\theta_W\alpha^B, \nonumber \\
\alpha^Z=&&-\sin\theta_W\alpha^B, \nonumber \\
\beta^\gamma=&&\sin\theta_W\beta^W+\cos\theta_W\beta^B, \nonumber \\
\beta^Z=&&\cos\theta_W\beta^W-\sin\theta_W\beta^B.
\end{eqnarray}

\noindent After these definitions, the Feynman rules including quartic vertex can be obtained as follows \cite{fer4,fer5}

\begin{eqnarray}
\label{v1}
\Gamma_{\gamma t\bar{q}}=&&\frac{1}{\Lambda^2}[\gamma_\mu\gamma_{R}(\alpha_{tj}p_2+\alpha^*_{jt}p_1)+\hat{v}\sigma_{\mu\nu}(\beta_{tj}\gamma_{R}+\beta^*_{jt}\gamma_L)]
(k^{\mu}g^{\nu\alpha}-k^{\nu}g^{\mu\alpha}), \nonumber \\
\Gamma_{\gamma \bar{t}q} =&&\frac{1}{\Lambda^2}[\gamma_\mu\gamma_{R}(\alpha_{tj}p_1+\alpha^*_{jt}p_2)+\hat{v}\sigma_{\mu\nu}
(\beta_{tj}\gamma_{R}+\beta^*_{jt}\gamma_L)](k^{\mu}g^{\nu\alpha}-k^{\nu}g^{\mu\alpha}), \nonumber \\
\Gamma_{\gamma\gamma t\bar{q}}=&&\frac{g_e}{\Lambda^2}[(\noindent \not{k_1}g_{\mu\nu}-k_{1\nu}\gamma_\mu)\gamma_R(\alpha_{jt}+\alpha^*_{tj})+({k_2}g_{\mu\nu}-k_{1\mu}\gamma_\nu)\gamma_R(\alpha_{jt}+\alpha^*_{tj})].
\end{eqnarray}

\noindent Here, $\sigma_{\mu\nu}=\frac{i}{2}[\gamma_\mu,\gamma_\nu]$, $\gamma_{L(R)}$ are the left(right)-handed projection operators, $\hat{v}=v/\sqrt{2}=174$ GeV, $g_e=\sqrt{4\pi\alpha}$, $k_1$ and $k_2$ are the photon momentums,  $p_1$ , $p_2$ are $t$ and $q=u,c$ quark momentums, respectively. In $\Gamma_{\gamma t\bar{q}}$ and $\Gamma_{\gamma\gamma t\bar{q}}$   $t$-quark ($q$-quark) is incoming (outgoing) the vertex, in $\Gamma_{\gamma\bar{t}q}$ $t$-quark ($q$-quark) is outgoing (incoming) the vertex. Additionally, the momentum of the photons are incoming to the vertex.

\section{Photon-Photon Interactions at the LHC}

The Large Hadron Collider (LHC) provides high energetic proton-proton collisions with high luminosity. Therefore, it generates very rich statistical data. It is expected that LHC will answer many unknown problems in new theories. However, ultraperipheral interactions and elastic collisions may not be catched at the main detectors of the LHC with limited pseudorapidity. For this reason, ATLAS and CMS Collaborations developed a plan of forward physics with updated extra detectors. These extra detectors are placed at a distance of $220$ m - $420$ m from the interaction point, in order to detect intact protons which are scattered after the collisions with some momentum fraction loss  $\xi=(|E|-|E^{\,\,\prime}|)/|E|$. Here $E$ is the energy of the incoming proton and $E^{\,\,\prime}$ is the energy of intact scattered proton. These new machines are known very forward detectors (VFDs). With VFDs, it will be possible to study the exclusive interactions of proton-proton and opens new opportunities of studying high energy photon-induced reactions, such as photon-photon and photon-proton interactions.
The $pp$ deep inelastic scattering (DIS) have very complex backgrounds due to interacted protons dissociate into partons. In the DIS process, made up of jets would cause some ambiguities. This situation make it hard to detect the new physics signals beyond the SM.
On the other hand, $\gamma\gamma$ or $\gamma p$ collisions have lower backgrounds than proton-proton DIS. Because, in photon induced reactions quasi-real photons emitted from proton beam can interact with other protons or emitted photons. The emitted almost-real photons have a low virtuality. Therefore, the proton structure remains intact. Moreover, $\gamma\gamma$ collisions are the most clean processes since they do not include any QCD interactions.

VFDs can detect intact outgoing protons in the interval $\xi_{min}<\xi<\xi_{max}$. This interval is known as the acceptance of the VFDs. If these machines are established closer to central detectors, a higher $\xi$ can be obtained. One of the programs about these detectors was prepared by ATLAS Forward Physics Collaboration (AFP). This program includes $0.0015 < \xi <0.15$, $0.015 < \xi <0.15$ detector acceptance ranges \cite{albrow}. It is organized to two types of measurements to research with high precision using the AFP \cite{afp1,afp2,afp3}. These are exploratory physics (anomalous couplings between
$\gamma$ and $Z$ or $W$ bosons, exclusive production, etc.) and
standard QCD physics (double Pomeron exchange, exclusive
production in the jet channel, single diffraction, $\gamma\gamma$ physics, etc.). These measurements will enhance the HERA and Tevatron
experiments to the kinematical region of the LHC. Furthermore, CMS-TOTEM forward detectors are placed closer to the central detectors and they have acceptance regions $0.0015 < \xi< 0.5$, $0.1< \xi <0.5$ \cite{lhc6, avati}. The main goals of the TOTEM
experiment are examining the elastic proton-proton interactions, total proton-proton cross-section,
and overall types of diffractive physical processes. The TOTEM experiment use the Roman Pots detector. It can be moved nearby to the outgoing protons to enable the trigger on elastic and diffractive protons
and to measure their physical parameters such as the momentum shift
and the transverse momentum exchange. Detectors of the
charged particle in the forward area can catch almost all
inelastic physical processes. A large solid
angle is covered with support of the CMS detector. Therefore, the detectors enable the workers to perform precise studies \cite{ttm1, ttm2, ttm3}. High energy scattering are accompanied by a number of soft interactions in the same bunch-crossing, known as pile-up events at high luminosity values. However, these backgrounds can be suppressed by using exclusivity conditions, kinematics and timing constraints at high luminosity values with application of forward detectors in conjunction with main detectors \cite{albrow, pl2, pl3}.

Photon-photon interactions were recently examined in the measurements of the CDF collaboration \cite{cdf1,cdf2,cdf3,cdf4,cdf5,cdf6,cdf7}. Their results are consistent in theoretical calculations with $p\bar{p} \to p\ell^{+}\ell^{-}\bar{p}$ through the subprocess ($\gamma\gamma \to \ell^{+}\ell^{-}$). At the LHC, the CMS collaboration have also detected photon-induced reactions $pp \to p\gamma \gamma p \to p\mu^+ \mu^- p $, $pp \to p\gamma \gamma p \to p e^+ e^- p $ from the $\sqrt{s}=7$ TeV \cite{ch1,ch2}. Therefore, the photon-induced interactions potential at the LHC is significant, with its high energy and high luminosity \cite{lhc1,lhc1a,lhc2a,lhc2,lhc4,lhc5,lhc7,khoze,albrow2,inanc,inan,kepka,bil,bil2,kok,inan2,ban,gru,inanc2,epl,inanc3,bil4,inanc4,hao1,hao2,
sen,kok2,ha1,ha2,tas,ins}.

\begin{figure}
\includegraphics{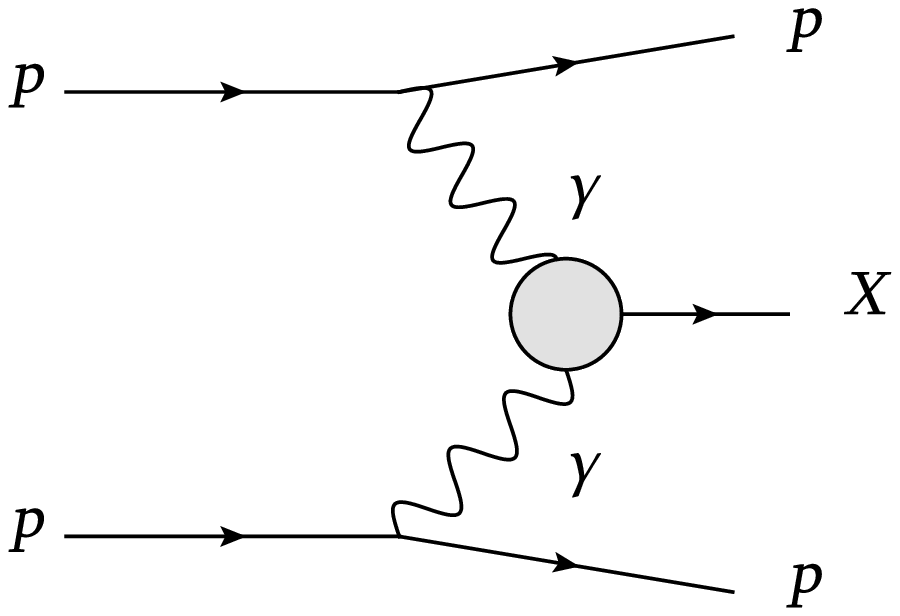}
\caption{Shematic diagram for the reaction $pp\to p \gamma \gamma p \to pXp$.}
\label{fig1}
\end{figure}

As mentioned above, forward detectors make it possible to
measure high energy photon-photon interaction. This process
is occurred by the collision of two photons which are radiated off the incoming protons and produce a central system $X$ through the process $pp \to p\gamma \gamma p \to pXp $. Schematic diagram for this process can be seen in Fig.\ref{fig1}. The system
X will be detected by the central detector under clean experimental conditions. Two
protons remain intact due to low virtuality of photons. These intact protons are also known as forward protons. They can not catched at the main detectors and
go on their path near to the beam line. Because energy loses of the protons can be measured by the forward detectors, it is possible
to know invariant mass of the central system $W=2E\sqrt{\xi_1 \xi_2}$.

At the LHC, the equivalent photon approximation (EPA) has been satisfyingly applied to photon-induced reactions \cite{ep,budnev,baur}.
In this method, two quasi-real photons with low virtuality are ($Q^2 = -q^2$) emitted
by each incoming proton. These photons interact with each other to produce X through the subprocess $\gamma\gamma \to X$.
The emitted quasi-real photons give a spectrum in terms of virtuality $Q^2$ and the photon energy $E_\gamma=\xi E$,

\begin{eqnarray}
\frac{dN}{dE_{\gamma}dQ^{2}}=\frac{\alpha}{\pi}\frac{1}{E_{\gamma}Q^{2}}
[(1-\frac{E_{\gamma}}{E})
(1-\frac{Q^{2}_{min}}{Q^{2}})F_{E}+\frac{E^{2}_{\gamma}}{2E^{2}}F_{M}]
\label{phs}
\end{eqnarray}

\noindent where $m_{p}$ is the mass of the proton.
The other terms are as follows,

\begin{eqnarray}
Q^{2}_{min}=\frac{m^{2}_{p}E^{2}_{\gamma}}{E(E-E_{\gamma})},
\;\;\;\; F_{E}=\frac{4m^{2}_{p}G^{2}_{E}+Q^{2}G^{2}_{M}}
{4m^{2}_{p}+Q^{2}} \\
G^{2}_{E}=\frac{G^{2}_{M}}{\mu^{2}_{p}}=(1+\frac{Q^{2}}{Q^{2}_{0}})^{-4},
\;\;\; F_{M}=G^{2}_{M}, \;\;\; Q^{2}_{0}=0.71 \mbox{GeV}^{2}.
\end{eqnarray}

\noindent Here, $F_{E}$ and $F_{M}$ are the functions of the electric and magnetic form factors respectively,  $\mu_{p}^2=7.78$ is the squared
magnetic moment of the proton. This spectrum differs from the pointlike electron case by taking into account of the electromagnetic form factors.
The luminosity spectrum of photon-photon collisions $\frac{dL^{\gamma\gamma}}{dW}$ can be obtained in the framework of the EPA as follows,

\begin{eqnarray}
\label{efflum}
\frac{dL^{\gamma\gamma}}{dW}=\int_{Q^{2}_{1,min}}^{Q^{2}_{max}}
{dQ^{2}_{1}}\int_{Q^{2}_{2,min}}^{Q^{2}_{max}}{dQ^{2}_{2}} \int_{y_{
min}}^{y_{max}} {dy \frac{W}{2y} f_{1}(\frac{W^{2}}{4y}, Q^{2}_{1})
f_{2}(y,Q^{2}_{2})}.
\end{eqnarray}

\noindent Here, we have taken the $Q_{max}^2=2$ GeV$^2$ since $Q_{max}^2$ is greater than $2$ GeV$^2$ region does not make a significant contribution to this integral. From Eq.(\ref{efflum}) the cross section for the main process $pp \to p \gamma \gamma p \to p X p $ can be found by integrating $\gamma \gamma \to X$ subprocess cross section over the photon spectrum,

\begin{eqnarray}
\label{completeprocess}
 d\sigma=\int{\frac{dL^{\gamma\gamma}}{dW}}
\,d\hat {{\sigma}}_{\gamma\gamma \to X}(W)\,dW.
\end{eqnarray}

\noindent In this paper, we have examined the anomalous FCNC interactions for the process $p p\to p\gamma\gamma p\to pt\bar{q}p$ at the LHC through the subprocess  $\gamma \gamma\to t\bar{q}$. In all results of this study, we impose a cut of $|\eta|<2.5$ and $p_t>30$ GeV. The QED two-photon
survival probability have been taken as $0.9$ \cite{koh}. Additionally, we have assumed that the center-of-mass energy of the LHC is
$14$ TeV.

\section{Numerical Analysis}

The effective operator methods provide to obtain the possible rare decays of
the top quark in a model-independent manner. The squared amplitude for top FCNC decay $t\to q\gamma$ ($q=u,c$) can be obtained in terms of the anomalous couplings by using Eq.(\ref{v1}),

\begin{eqnarray}
\label{mbr}
|M_{t\to \gamma q}|^{2}=\frac{m_t^4}{2\Lambda^4}\{{m_t^2|\alpha_{tu}^\gamma+(\alpha_{ut}^\gamma)^*|^2+16\hat{v}^2(|\beta_{tu}^\gamma)|^2+|\beta_{ut}^\gamma)|^2)+
8\hat{v}m_t \textrm{Im}[\beta_{tu}^\gamma(\alpha_{ut}^\gamma+(\alpha_{ut}^\gamma)^*)]}\}. \nonumber \\
\end{eqnarray}

\noindent From this result, it is easy the obtain decay width,

\begin{eqnarray}
\label{drt}
\Gamma_{t\to \gamma q}=\frac{m_t^3}{64\pi\Lambda^4}\{m_t^2|\alpha_{tu}^\gamma+(\alpha_{ut}^\gamma)^*|^2+16\hat{v}^2(|\beta_{tu}^\gamma)|^2+|\beta_{ut}^\gamma)|^2)+
8\hat{v}m_t \textrm{Im}[\beta_{tu}^\gamma(\alpha_{ut}^\gamma+(\alpha_{ut}^\gamma)^*)]\}. \nonumber \\
\end{eqnarray}

\begin{figure}
\centering
\includegraphics{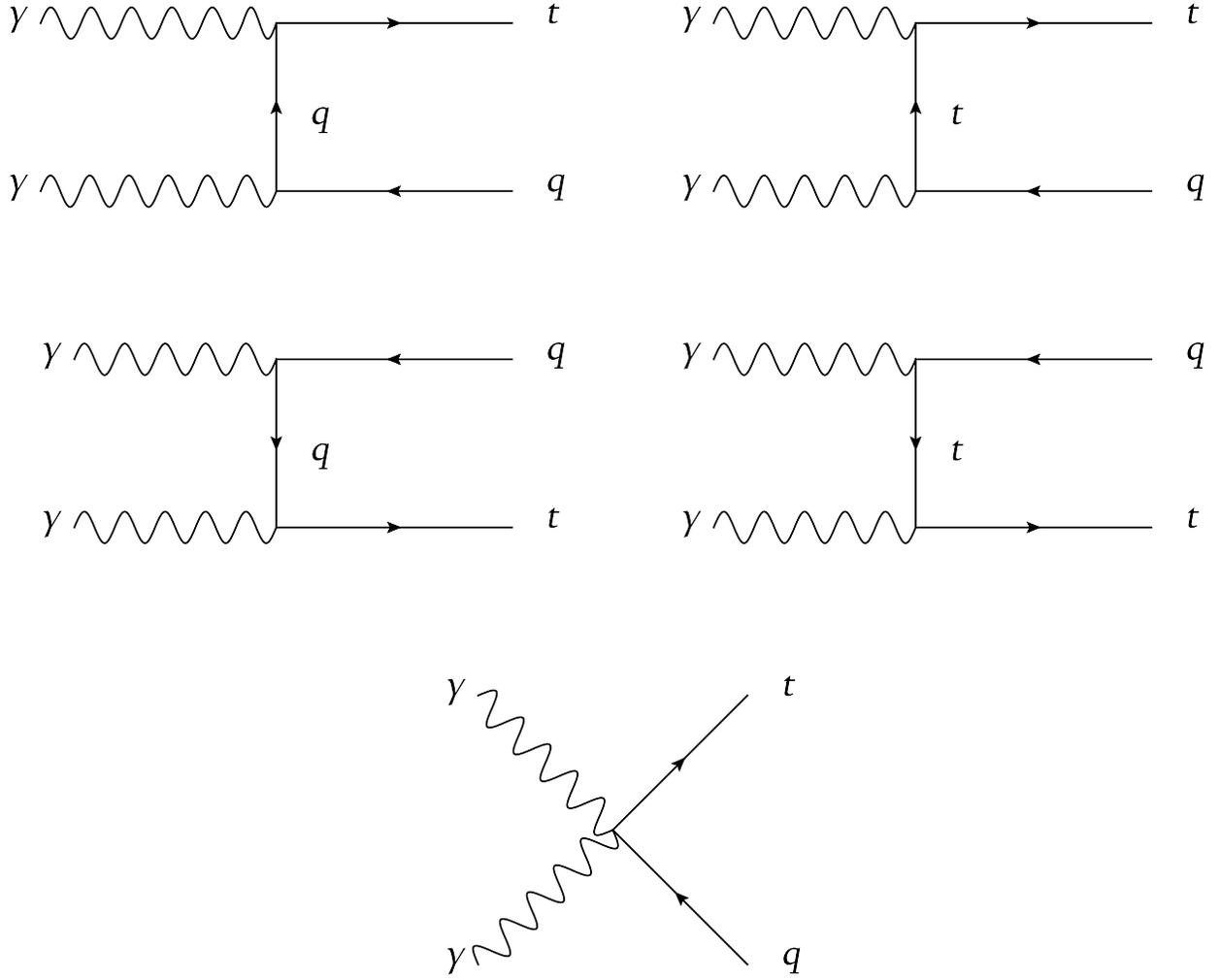}
\caption{Three level Feynman diagrams for the subprocess $\gamma \gamma \to t\bar{q}$ ($q=u,c$) in the presence of the anomalous dimension-six $tq \gamma$ couplings.}
\label{fig2}
\end{figure}

\noindent There are five Feynman Diyagrams for the $\gamma\gamma\to t\bar{q}$ as shown in Fig.\ref{fig2}. The polarization summed amplitude square can be found by using Eq.(\ref{v1}),

\begin{eqnarray}
\label{csa}
|M_{\gamma\gamma\to t\bar{q}}|^{2}&&=\frac{g_e^2Q_t^2s}{\Lambda^4(t-m_t^2)^2t(u-m_t^2)^2u}\{m_t^{10}(t+u)-12m_t^8tu+m_t^6(t+u)(t^2+13tu+u^2) \nonumber \\
&&-m_t^4tu(t^2+24tu+7u^2)+12m_t^2t^2u^2(t+u)-6t^3u^3\}\{m_t^2|\alpha_{tu}^\gamma+(\alpha_{ut}^\gamma)^*|^2 \nonumber \\
&&+16\hat{v}^2(|\beta_{tu}^\gamma)|^2+|\beta_{ut}^\gamma)|^2)+8\hat{v}m_t \textrm{Im}[\beta_{tu}^\gamma(\alpha_{ut}^\gamma+(\alpha_{ut}^\gamma)^*)]\}.
\end{eqnarray}

\noindent where $s=(p_1+p_2)^2=(k_1+k_2)^2, t=(k_1-p_1)^2=(k_2-p_2)^2$ and $u=(k_1-p_2)^2=(k_2-p_1)^2$ are the Mandelstam variables. The differential cross section for the $\gamma\gamma\to t\bar{q}$ can be written by the means of the decay rate as seen from the Eqs.(\ref{drt}) and (\ref{csa}),

\begin{eqnarray}
\label{dcs}
\frac{d\sigma}{d(\cos\theta)}=\frac{3(s-m_t^2)Q_t^2g_e^2}{2m_t^3stu(t-m_t^2)^2(u-m_t^2)^2}G_{\gamma\gamma}\Gamma_{t\to q \gamma}.
\end{eqnarray}

\noindent Here $G_{\gamma\gamma}$ function given as follows,

\begin{eqnarray}
G_{\gamma \gamma}&&=m_t^{10}(t+u)-12m_t^8tu+m_t^6(t+u)(t^2+13tu+u^2) \nonumber \\
&&-m_t^4tu(t^2+24tu+7u^2)+12m_t^2t^2u^2(t+u)-6t^3u^3.
\end{eqnarray}

\begin{figure}
\includegraphics{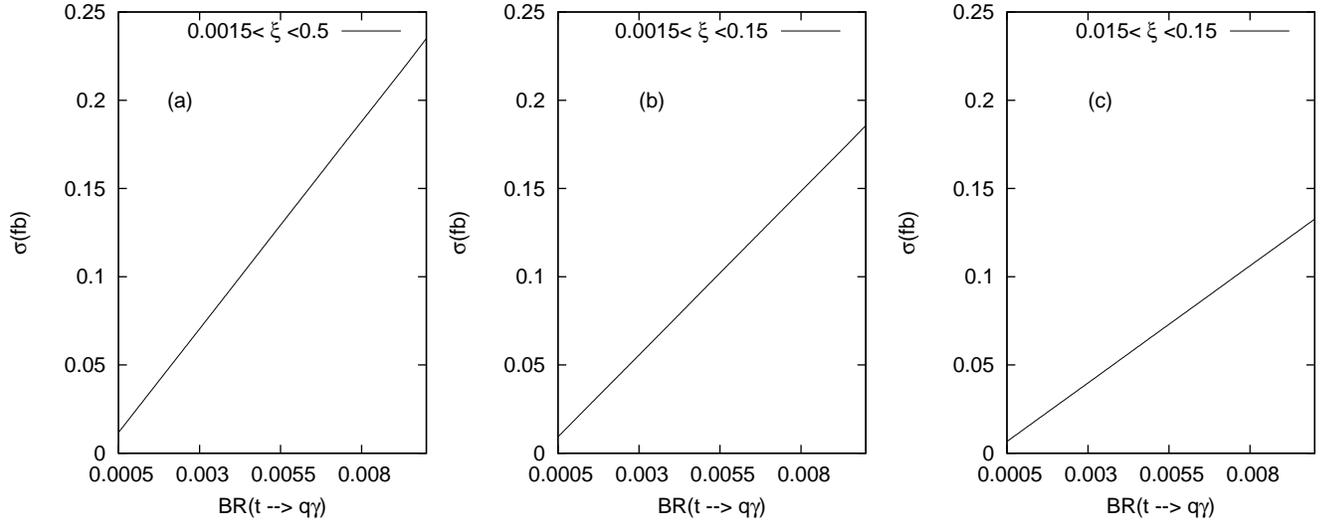}
\caption{The cross sections of $pp \to p\gamma \gamma p \to p t\bar{q} p$ as a function of branching ratios of $t\to q\gamma$ ($BR(t \to q\gamma$)) for three forward detector acceptances: $0.0015< \xi < 0.5$, $0.0015< \xi < 0.15$ and $0.015< \xi <0.15$.}
\label{fig3}
\end{figure}

\begin{figure}
\includegraphics{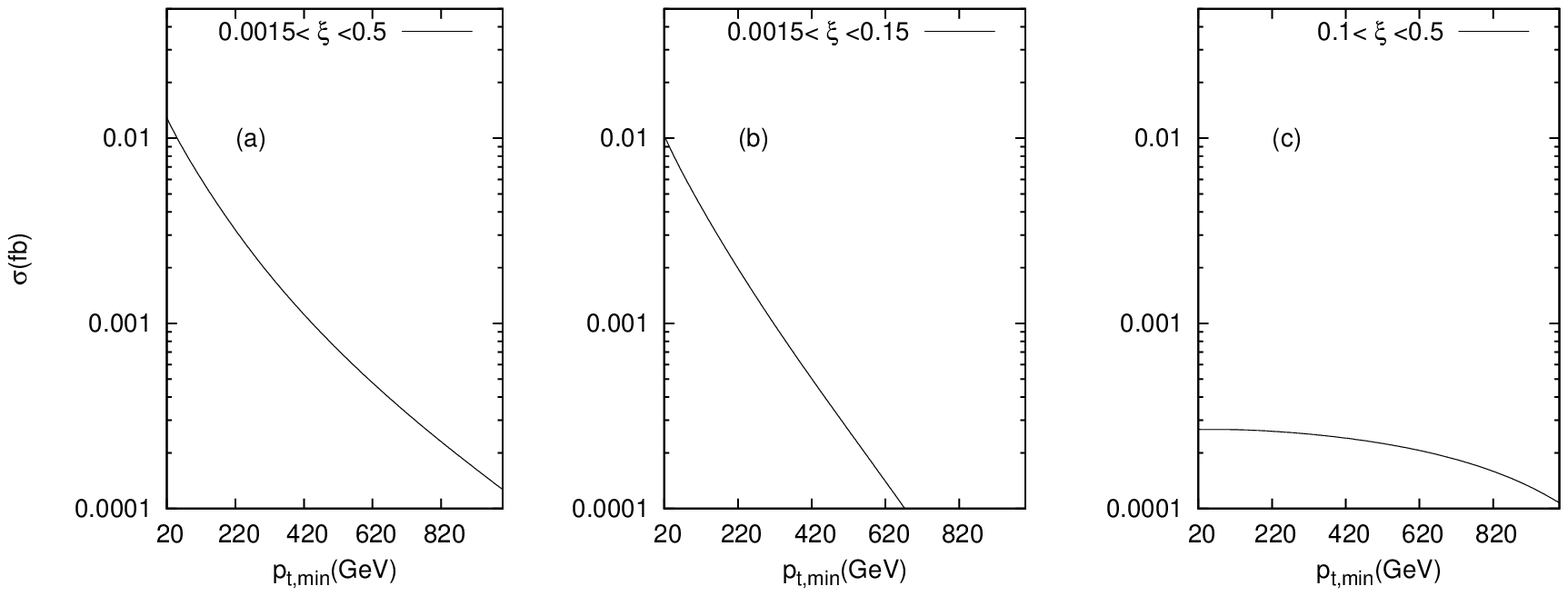}
\caption{Cross sections of $pp \to p\gamma \gamma p \to pt\bar{q}p$  as a function of
the transverse momentum cut on the final state particles for $BR(t\to q\gamma)=0.0005$ and three forward detector acceptances: $0.0015< \xi <0.5$, $0.0015< \xi <0.15$,
and $0.1< \xi <0.5$.}
\label{fig4}
\end{figure}

\begin{figure}
\includegraphics{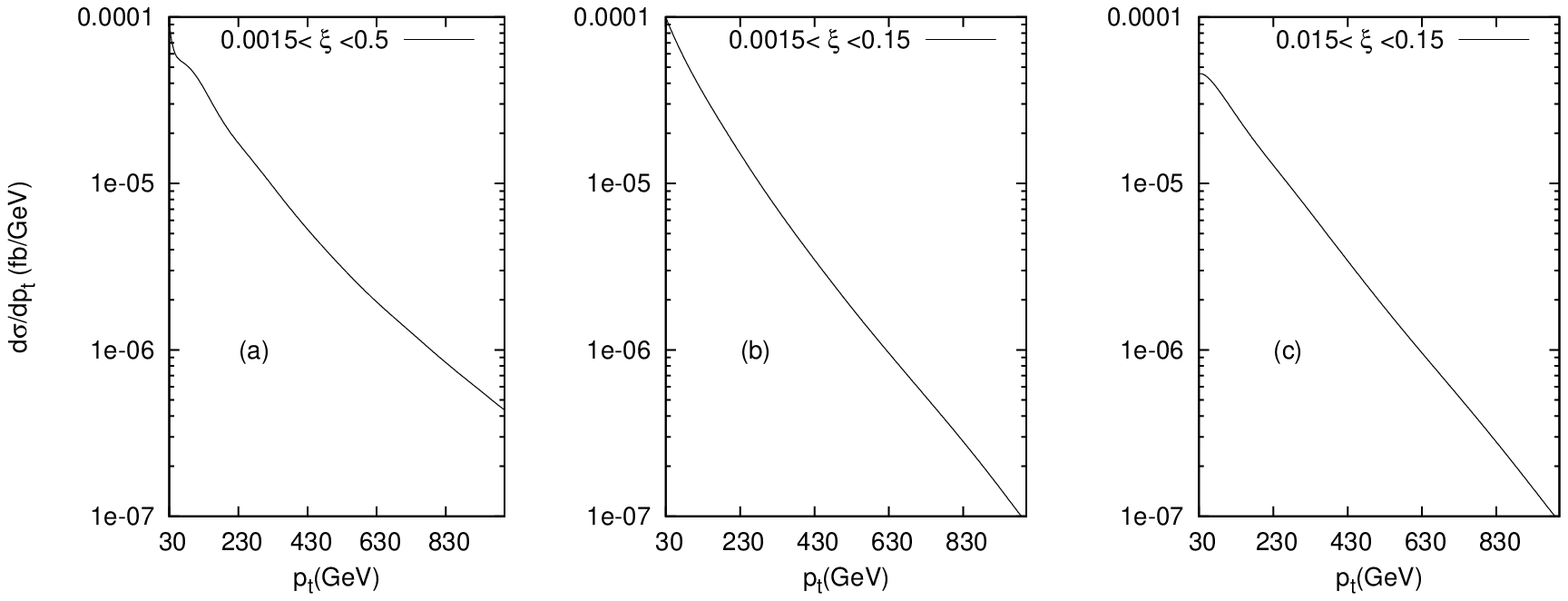}
\caption{Differential cross sections of $pp \to p\gamma \gamma p \to pt \bar{q} p$  as a function of
the transverse momentum on the final state particles for $BR(t\to q\gamma)=0.0005$ and three forward detector acceptances: $0.0015< \xi <0.5$, $0.0015< \xi <0.15$,
and $0.015< \xi <0.15$.}
\label{fig5}
\end{figure}

\noindent In figures \ref{fig3}(a-c), we show the total cross sections as functions of branching ratio of the ${t\to q\gamma}$ decay for three acceptance regions:  $0.0015< \xi < 0.5$, $0.0015< \xi <0.15$ and $0.015< \xi < 0.15$. We obtain from these figures that the total cross section for the $0.0015< \xi < 0.5$ is better than the others. Also, we have calculated the cross section for the $0.1< \xi < 0.5$ acceptance range. However, the cross section for the this acceptance range is very small. For instance, it has been obtained $2.75\times10^{-4}$ fb for $BR(t\to q\gamma)=0.0005$. Hence, we do not show the cross section for this acceptance range. This result can be understood from the figures \ref{fig4}(a-c). These figures represent the cross sections versus the minimum transverse momenta (or $p_{t cut}$) of the final quarks for $BR(t\to q\gamma)=0.0005$. When Figs. \ref{fig4}(a) and \ref{fig4}(c) are compared, it can be obtained that the acceptance region $0.1< \xi < 0.5$ has almost the same result as the region $0.0015< \xi < 0.5$ with $p_{t,min}=800$ GeV. Therefore, the cross section with a high acceptance region's lower bound is similar to that with an additional $p_t$ cut. In figures \ref{fig5} (a-c), we plot the $p_t$ distribution of the final state quarks for differential cross section with  $BR(t\to q\gamma)=0.0005$ for three acceptance regions:  $0.0015< \xi < 0.5$, $0.0015< \xi <0.15$ and $0.015< \xi < 0.15$. It turns out that anomalous coupling has the dominant effect in low $p_t$ regions. Hence, $0.1< \xi < 0.5$ forward detector acceptance range is not convenient for investigating dimension-six anomalous top quark coupling.

It can be considered that, there are SM backgrounds. $pp\to p\gamma\gamma p \to pWb\bar{q}p$ process is one of the these backgrounds. However, this background is very small ($= 4.5\times10^{-6}$ fb) even for the $0.0015 < \xi < 0.5$ and therefore we do not consider this background. The process of $\gamma\gamma\to 4j$
in SM would contribute to this background if one of the light jet is mistaken to be a $b$-jet. We have found this background in order of $10^{-2}$ fb for $0.0015<\xi<0.5$. However, the new developments were reported in reducing the light quark-b misidentification probabilities in ATLAS \cite{atsb} and CMS \cite{cmsb}. In CMS experiment, a misidentification probability of only in order of $1\%$ has been achieved. The cross section of the signal for the  $pp\to p\gamma\gamma p \to p t \bar{q} p$  is in order of $10$ fb for the $0.0015<\xi<0.5$. Therefore, we think that the inclusion of these backgrounds to this paper may be neglected.

We have found $95\%$ confidence level (C.L.) limits on the branching ratios of the
top quark. We have applied the Poisson distribution statical analysis method since the SM background for the this process is absent.
In Poisson analysis, the number of observed events are assumed to be equal to the SM prediction. Upper bounds of
events number $N_{up}$ can be obtained from the following equation at the $95\%$ C.L. \cite{fav,pie},

\begin{eqnarray}
\sum_{k=0}^{N_{obs}}P_{Poisson}(N_{up},k)=0.05.
\end{eqnarray}

\begin{figure}
\includegraphics{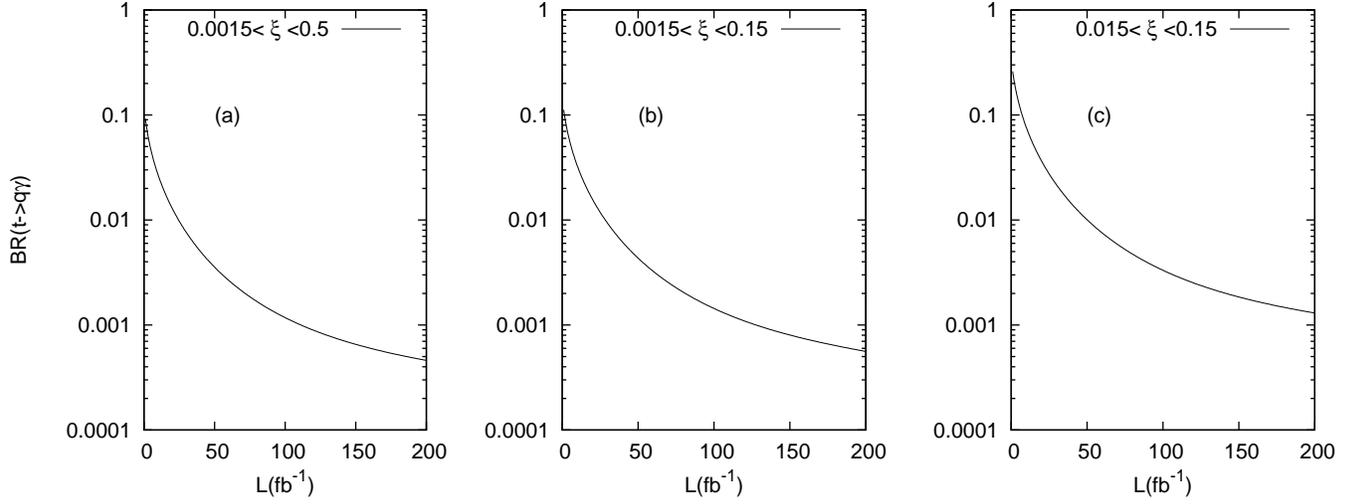}
\caption{$95\%$ C.L. lower bounds for branching ratios of $t\to q \gamma$ ($BR(t \to q \gamma$)) as a function of integrated LHC luminosity for three forward detector acceptances: $0.0015< \xi <0.5$, $0.0015< \xi <0.15$ and $0.015< \xi <0.15$.}
\label{fig6}
\end{figure}

\noindent Depending on the number of observed events, values for upper limits $N_{up}$ can be found in Table $38.3$ in Ref.\cite{par}.
Since $N_{obs}=0$ in our paper, we have taken $N_{up}=3$.  The $N_{up}$ can be directly converted to the bound of branching ratio of $t\to q\gamma$ with using Eq.(\ref{dcs}) for the different luminosity values.
In Figs.\ref{fig6}(a-c), we represent sensitivity bounds on the $BR(t\to q\gamma)$. These bounds are given as a function of
integrated LHC luminosity for three forward detector acceptances $0.0015 < \xi < 0.5$, $0.0015 < \xi < 0.15$ and $0.015 < \xi< 0.15$. We see from these figures that our limits are better than the current experimental best stringent result for $t\to c\gamma$. At the same time, even at the next searches of the LHC $pp$ collisions with $3000$ $fb^{-1}$ of integrated luminosity, LHC sensitivity bounds on $BR(t\to q\gamma)$ would not be improved substantially \cite{cmss,ats}. Therefore, it may be important to examine FCNC anomalous coupling of the top quark at future photon-induced LHC studies with very high luminosity values.

On the other hand, FCNC Lagrangian considered in \cite{sa1,sa2} to define dimension-six anomalous interaction contains two effective operators instead of four ones. It has been showed that the operator $O_{tb}$ in Eq.\ref{oprs} is redundant. Then, the author have obtained the following interaction lagrangian,

\begin{eqnarray}
\textit{L}_{\gamma tq}=-g_e\bar{q}\frac{i\sigma^{\mu\nu}q_\nu}{m_t}(\lambda^L\gamma_L+\lambda^R\gamma_R)tA_\mu+H.c..
\label{lag2}
\end{eqnarray}

\noindent This lagrangian includes the same physics, under change of variables plus some redefinitions of for fermion operator coefficients. With using this effective lagrangian, decay width for $t\to q\gamma$ can obtain much simpler form,

\begin{eqnarray}
\Gamma(t\to q\gamma)=\frac{g_e^2mt}{16\pi}(|\lambda^R|^2+|\lambda^L|^2).
\end{eqnarray}

\begin{figure}
\includegraphics{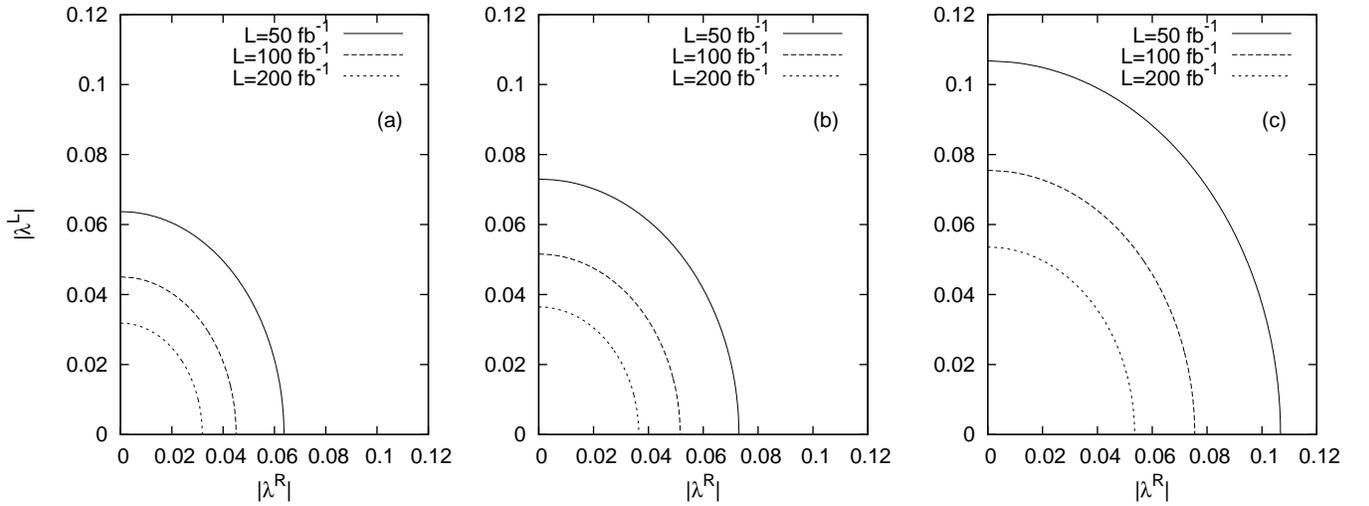}
\caption{$\%95$ C.L. contours for $|\lambda^R|$-$|\lambda^L|$ for $L=50 fb^{-1}$, $L=100 fb^{-1}$, $L=200 fb^{-1}$ and
three forward detectors acceptance regions: $0.0015 < \xi < 0.5$, $0.0015 < \xi < 0.15$ and $0.015 < \xi< 0.15$.}
\label{fig7}
\end{figure}

\noindent The differential cross section is also same as Eq.(\ref{dcs}). Therefore, our discussion do not change for this effective lagrangian. Additionally, we have obtained $\%95$ C.L. contours for $\lambda^R$ and $\lambda^L$ for $L=50 fb^{-1}$, $L=100 fb^{-1}$, $L=200 fb^{-1}$ and
three forward detectors acceptance regions  $0.0015 < \xi < 0.5$, $0.0015 < \xi < 0.15$ and $0.015 < \xi< 0.15$ in the Figs.\ref{fig7}(a-c).

Furthermore, we have calculated spin asymmetry of the final state single top quark with using Eq.(\ref{lag2}). The correlation among the top spin and its decay products can be obtained in the rest frame of the final state top quark. In this situation, the angular distribution of the decay is obtained as follows,

\begin{eqnarray}
\label{asm}
\frac{1}{\Gamma_T}\frac{d\Gamma}{d\cos\theta_i}=\frac{1}{2}(1+\alpha_i\cos\theta_i),
\end{eqnarray}

\noindent where, ${\Gamma_T}$ is the total decay rate of the top quark, $\theta_i$ is the angle between the decay product and the top quark spin quantization axis and $\alpha_i$ is the correlation
degree between the decay products and top spin ($\alpha_i = 1$ for $i =l^+, \bar{d}, \bar{s}$; $\alpha_i = 0.4$ for $i=b$) \cite{ma}. If there is a mixture of spin up and spin down top quarks in the interaction, the Eq.(\ref{asm}) turns into following form,

\begin{eqnarray}
\label{asmm}
\frac{1}{\Gamma_T}\frac{d\Gamma}{d\cos\theta_i}=\frac{1}{2}(1+A\alpha_i\cos\theta_i).
\end{eqnarray}

\noindent Here, A is called the spin asymmetry. In order to find the cross section, depending on the spin, the following projection operator can be used,

\begin{eqnarray}
\sum_{s_t}u(p_1,s_t)\bar{u}(p_1,s_t)=\frac{1}{2}(1+\gamma_5 \not{s_t})(\not{p_1}+m_t)
\end{eqnarray}

\noindent where $s_t$ is the spin vector of the top quark. It can be established in the helicity basis as follows,

\begin{eqnarray}
s_t^\mu=\lambda_t\left(\frac{|\vec{p_1}|}{m_t},\frac{E_1}{m_t}\frac{\vec{p_1}}{|\vec{p_1}|}\right); && \lambda_t=\pm1,
\end{eqnarray}

\noindent here, $E_1$ is the energy of the top quark. In this case, spin asymmetry of the top quark can be written in terms of spin dependent events number $N(\lambda_t)$ as following form,

\begin{eqnarray}
A=\frac{N(\lambda_t=1)-N(\lambda_t=-1)}{N(\lambda_t=1)+N(\lambda_t=-1)}.
\end{eqnarray}

\noindent Depending on the helicity of the top quark $\lambda_t$, differential cross section can be obtained for $\gamma\gamma\to t\bar{q}$ subprocess,

\begin{eqnarray}
\label{pcs}
\frac{d\sigma(\lambda_t)}{d(\cos\theta)}&&=-\frac{3g_e^4Q_t^2(s-mt^2)}{128\pi mt^2s^2tu(t-mt^2)^2(u-mt^2)^2}[(|\lambda^{R}|^2+|\lambda^L|^2)2sG_{\gamma\gamma} \nonumber \\
&&+\lambda_t(|\lambda^{R}|^2-|\lambda^{L}|^2)H_{\gamma\gamma}],
\label{pp}
\end{eqnarray}

\noindent where,

\begin{eqnarray}
H_{\gamma\gamma}&&=\frac{1}{t+u}[(16tu)mt^{12}-4(t+u)(t^2+u^2+12tu)mt^{10} \nonumber \\
&&+(76tu^3+76ut^3+176t^2u^2)mt^8 \nonumber \\
&&+(t+u)(2t^4+2u^4-47tu^3-47ut^3-166t^2u^4)mt^6 \nonumber \\
&&+(9tu^5+9ut^5+108t^2u^4+108u^4t^2+222t^3u^3)mt^4 \nonumber \\
&&-(t+u)(19t^2u^4+19u^4t^2+74t^3u^3)mt^2+12t^3u^3(t+u)^2].
\end{eqnarray}

\begin{figure}
\includegraphics{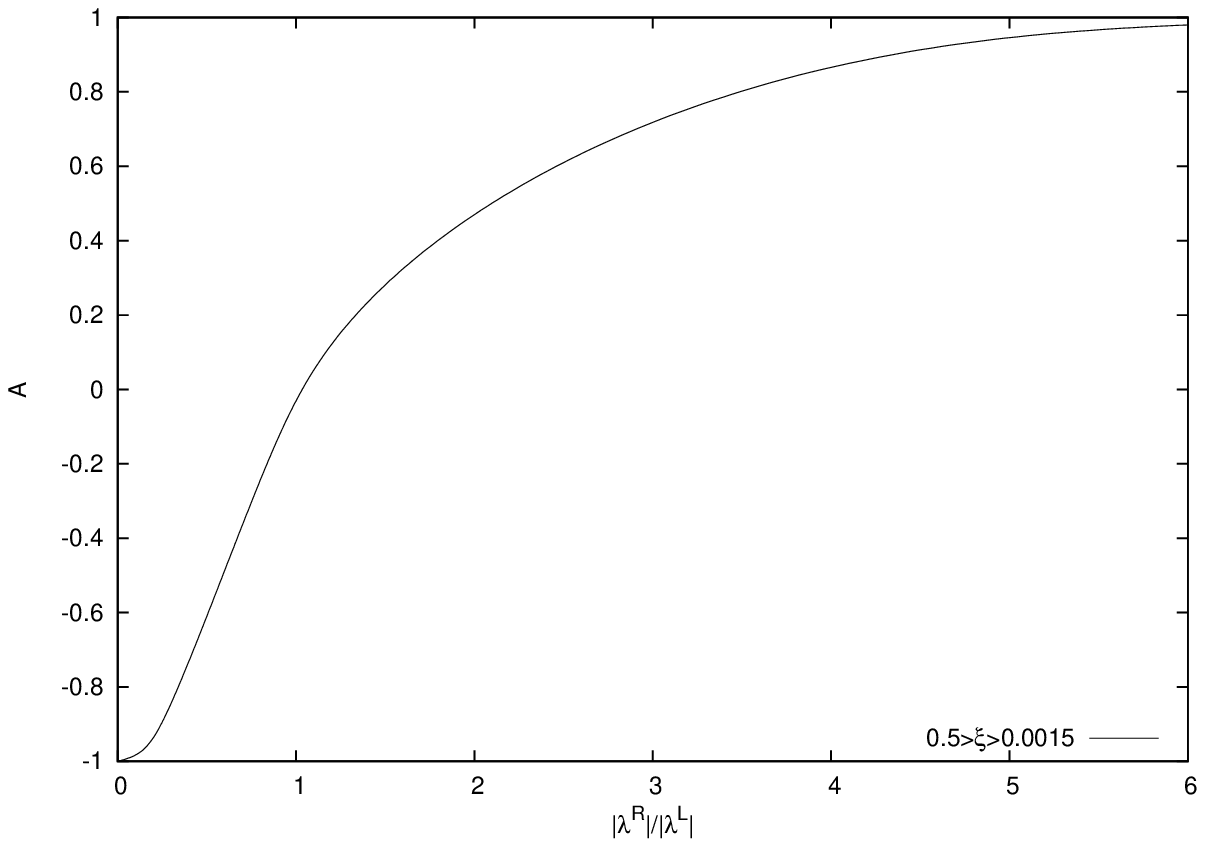}
\caption{Single top quark spin asymmetry as function of the $|\lambda^R|$/$|\lambda^L|$ for $0.0015 < \xi < 0.5$.}
\label{fig8}
\end{figure}

\noindent There is no polarization of the top quark when $\frac{|\lambda_R|}{|\lambda_L|}=1$ as showed from the Eq. \ref{pp}. In Fig.\ref{fig8}, we have plotted the top quark spin asymmetry as function of the $\frac{|\lambda_R|}{|\lambda_L|}$ for $0.0015< \xi < 0.5$. We have also obtained the spin asymmetry for $0.0015< \xi < 0.15$ and $0.015< \xi < 0.15$. However, since these results very similar to $0.0015< \xi < 0.5$ case, we do not show the figures for these acceptance ranges. As seen from the Fig.\ref{fig8}, when $\frac{|\lambda_R|}{|\lambda_L|}$ goes to $0$ (infinity), asymmetry approach the $-1$ ($1$). Therefore, asymmetry can be used to determine the type of the interaction lagrangian.

\section{Conclusions}

The LHC can be used as a high energy photon-photon and photon-proton collider with new equipments which are called very forward detectors. There are no existing high energy photon-photon, photon-proton colliders with this quality. Particle production through $\gamma\gamma$ fusion yield fewer backgrounds than the pure DIS process. There are no proton remnants after the collisions and, these type of interactions are only electromagnetic in nature. The intact protons detect in forward detectors. This detection allow to measure the energy of the almost-real photons. In this case, it is possible to determine the invariant mass of the central system. With this clean environment, any discrepant signal with the prospect of the SM would be a conclusive clue for new physics beyond the SM. Moreover, anomalous $tq\gamma$ couplings might also be uniquely revealed in single top photon-induced reactions \cite{albrow}.

In this paper, we have examined anomalous dimension-six top quark
photon couplings in a model-independent way in the $pp\to p\gamma\gamma p \to p t\bar{q}p$ process for three forward detector acceptances $0.0015< \xi < 0.5$, $0.0015< \xi < 0.15$ and, $0.015< \xi <0.15$.  We have obtained the sensitivity bounds on branching ratio of the $t\to q\gamma$ and anomalous couplings. We see that, our obtained results can improve the sensitivity bounds for the branching ratio of the $t \to c\gamma$  with
respect to current experimental results. We have also made these analysis for another dimension-six effective lagrangian which have only two anomalous couplings. This effective operator contains the same physics. Therefore, we have obtained the same results for the cross sections and sensitivity bounds. Additionally, we have analyzed spin asymmetry of the single top quark through the process $pp\to p\gamma\gamma p \to pt\bar{q}p$ for this effective lagrangian. We have seen that the asymmetry is very sensitive to the couplings. Hence, asymmetry can be used in determining the structure of the interaction lagrangian. Based on the findings of this study, it is concluded that, photon-photon fusion provides new opportunities for top quark physics beyond the SM.

\pagebreak

\pagebreak

\begin{thebibliography}{99}

\bibitem{ben} M. Beneke et al., arXiv:0003033.
\bibitem{cha} D. Chakraborty, J.
Konigsberg, and D. Rainwater, Annu. Rev. Nucl. Part.
Sci. 53, 301 (2003).
\bibitem{wag} W. Wagner, Rep. Prog. Phys. 68,2409 (2005).

\bibitem{agu} J. A. Aguilar-Saavedra, Nucl. Phys. B {\bf 812}, 181 (2009).

\bibitem{he} B. Grzadkowski, J. F. Gunion and P. Krawczyk, Phys. Lett. B {\bf 268} 106 (1991).
\bibitem{ee1} G. Eilam, J. L. Hewett and A. Soni, Phys. Rev. D {\bf 44} 1473 (1991).
\bibitem{ee2} G. Couture, C. Hamzaoui and H. Kønig, Phys. Rev. D {\bf 52} 1713 (1995).
\bibitem{8} J. A. Aguilar-Saavedra and B. M. Nobre, Phys. Lett. B {\bf 553} 251(2003).

\bibitem{9} F. del Aguila, J. A. Aguilar-Saavedra and R. Miquel, Phys. Rev. Lett. {\bf 82} 1628 (1999).

\bibitem{10} J. A. Aguilar-Saavedra, Phys. Rev. D {\bf 67} 035003 (2003). Erratum-ibid. D {\bf 69} 099901 (2004).

\bibitem{11} T. P. Cheng and M. Sher, Phys. Rev. D {\bf 35} 3484 (1987).

\bibitem{12} B. Grzadkowski, J. F. Gunion and P. Krawczyk, Phys. Lett. B {\bf 268} 106 (1991).

\bibitem{13} M. E. Luke and M. J. Savage, Phys. Lett. B {\bf 307} 387 (1993).

\bibitem{14} D. Atwood, L. Reina and A. Soni, Phys. Rev. D {\bf 53} 1199 (1996).

\bibitem{15} D. Atwood, L. Reina and A. Soni, Phys. Rev. D {\bf 55} 3156 (1997).

\bibitem{16} S. Bejar, J. Guasch and J. Sola, Nucl. Phys. B {\bf 600} 21 (2001).

\bibitem{17} C. S. Li, R. J. Oakes and J. M. Yang, Phys. Rev. D {\bf 49} 293 (1994). Erratum-ibid.D {\bf 56}:3156,(1997).

\bibitem{18} G. M. de Divitiis, R. Petronzio and L. Silvestrini, Nucl. Phys. B {\bf 504} 45 (1997).

\bibitem{19} J. L. Lopez, D. V. Nanopoulos and R. Rangarajan, Phys. Rev. D {\bf 56} 3100 (1997).

\bibitem{20} J. Guasch and J. Sola, Nucl. Phys. B {\bf 562} 3 (1999).

\bibitem{21} D. Delepine and S. Khalil, Phys. Lett. B {\bf 599} 62 (2004).

\bibitem{22} J. J. Liu, C. S. Li, L. L. Yang and L. G. Jin, Phys. Lett. B {\bf 599} 92 (2004).

\bibitem{23} J. J. Cao et al., Phys. Rev. D {\bf 75} 075021 (2007).

\bibitem{24} J. M. Yang, B.-L. Young and X. Zhang, Phys. Rev. D {\bf 58} 055001 (1998).

\bibitem{25} G. Lu, F. Yin, X. Wang and L. Wan, Phys. Rev. D {\bf 68} 015002 (2003).

\bibitem{26} G. P. K. Agashe, G. Perez and A. Soni, Phys. Rev.D {\bf 71} 016002 (2005).

\bibitem{27} G. P. K. Agashe and A. Soni, Phys. Rev. D {\bf 75} 015002 (2007).

\bibitem{cdf} F. Abe et al., (CDF Collaboration), Phys. Rev. Lett. {\bf 80} 2525 (1998).

\bibitem{zeus} ZEUS Collaboration, Phys. Lett. B, {\bf 708} 27-36 (2012).

\bibitem{he9} J. Carvalho et al., (ATLAS Collaboration), Eur. Phys. J. C {\bf 52}
999-1019 (2007).
\bibitem{he10} L. Benucci, A. Kyriakis, Nucl. Phys. Proc. Suppl. 177-178 (2008) 258-260.
\bibitem{he11} Efe Yazgan et al.,arXiv:1312.5435.
\bibitem{cms} CMS Collaboration, CMS-PAS-TOP-14-003
\bibitem{crt} S. Khatibi and M. Mohammadi Najafabadi, Phys. Rev. D {\bf 89} 054011 (2014).
\bibitem{bah}  W. Buchm\"{u}ller and D. Wyler, Nucl. Phys. B {\bf 268}, 621 (1986).
\bibitem{fer1}P. M. Ferreira, O. Oliveira, and R. Santos, Phys. Rev. D {\bf 73},
034011 (2006).
\bibitem{fer2}P. M. Ferreira and R. Santos, Phys. Rev. D {\bf 73}, 054025
(2006).
\bibitem{fer3}P. M. Ferreira and R. Santos, Phys. Rev. D {\bf 74}, 014006
(2006).
\bibitem{fer4} P. M. Ferreira, R. B. Guedes and R. Santos, Phys. Rev. D {\bf 77} 114008 (2008).
\bibitem{fer5} R. A. Coimbra {\it et al.}, Phys. Rev. D {\bf 79} 014006 (2009).



\bibitem{albrow} M. Albrow {\it et al.},(FP420 R and D Collaboration), JINST 4, T10001 (2009).
\bibitem{afp1} ATLAS Collaboration,  Tech. Rep. CERN-LHCC-2011-
012. LHCC-I-020, (2011).
\bibitem{afp2} L. Adamczyk {\it et al.}, Tech. Rep. ATLCOM-
LUM-2011-006, CERN, 2011.
\bibitem{afp3} O. Kepka {\it et al.}, Tech.
Rep. ATL-COM-PHYS-2012-775, CERN, (2011).
\bibitem{lhc6} O. Kepka and C. Royon, Phys. Rev. D {\bf 78}, 073005 (2008).
\bibitem{avati} V. Avati and K. Osterberg, Report No. CERN-TOTEM-NOTE-2005-002, (2006).
\bibitem{ttm1} G. Antchev {\it et al.} (The TOTEM Collaboration),  EPL 96 21002, (2011).
\bibitem{ttm2} G. Antchev {\it et al.} (The TOTEM Collaboration),  EPL 98 31002, (2012).
\bibitem{ttm3} G. Antchev {\it et al.} (The TOTEM Collaboration), CERNPH-EP-2012-239, (2012).
\bibitem{pl2} M. Tasevsky, Nucl. Phys. Proc. Suppl. 179-180, 187 (2008).
\bibitem{pl3} M. G. Albrow, T. D. Coughlin and J. R. Forshaw, Prog. Part. Nucl. Phys. {\bf 65}, 149 (2010)




\bibitem{cdf1} A. Abulencia {\it et al.}, (CDF Collaboration), Phys. Rev. Lett. {\bf 98}, 112001
(2007).

\bibitem{cdf2} T. Aaltonen {\it et al.}, (CDF
Collaboration), Phys. Rev. Lett. {\bf 99}, 242002 (2007).

\bibitem{cdf3} T. Aaltonen {\it et al.}, (CDF Run II Collaboration),
Phys. Rev. D {\bf 77}, 052004 (2008).

\bibitem{cdf4} T. Aaltonen {\it et al.}, (CDF Collaboration), Phys. Rev. Lett. {\bf 102}, 242001
(2009).

\bibitem{cdf5} T. Aaltonen {\it et al.}, (CDF Collaboration), Phys. Rev. Lett. {\bf 102}, 222002
(2009).

\bibitem{cdf6}  O. Kepka and C. Royon, Phys. Rev. D {\bf 76}, 034012
(2007).

\bibitem{cdf7} M. Rangel, C. Royon, G. Alves, J. Barreto
and R. Peschanski,  Nucl. Phys. B {\bf 774}, 53 (2007).

\bibitem{ch1} S. Chatrchyan {\it et al.}, (CMS Collaboration), JHEP {\bf1201}, 052 (2012).
\bibitem{ch2} S. Chatrchyan {\it et al.}, (CMS Collaboration), JHEP {\bf1211}, 080 (2012).

\bibitem{khoze}V. Khoze, A. Martin, R. Orava and M. Ryskin, Eur. Phys. J. {\bf C19}, 313 (2001).
\bibitem{albrow2} M. Albrow, T. Coughlin and J. Forshaw, Prog. Part. Nucl. Phys. {\bf65}, 149 (2010).
\bibitem{lhc1} I. F. Ginzburg and A. Schiller, Phys. Rev. D {\bf 57}, R6599 (1998).
\bibitem{lhc1a} I. F. Ginzburg and A. Schiller, Phys. Rev. D {\bf 60}, 075016 (1999).
\bibitem{lhc2a} S. Lietti, A. Natale, C. Roldao and R.
Rosenfeld, Phys. Lett. B {\bf 497}, 243 (2001).
\bibitem{lhc2} K. Piotrzkowski, Phys. Rev. D {\bf 63}, 071502(R) (2001).
\bibitem{lhc4} V. Goncalves and M. Machado, Phys. Rev. D {\bf
75}, 031502(R) (2007).
\bibitem{lhc5} M.V.T. Machado, Phys. Rev. D {\bf 78}, 034016
(2008).
\bibitem{lhc7} S. Ata\u{g}, S. C. \.{I}nan and \.{I}. \c{S}ahin, Phys. Rev. D {\bf 80}, 075009 (2009).
\bibitem {inanc} \.{I}. \c{S}ahin and S. C. \.{I}nan, JHEP {\bf09}, 069 (2009).
\bibitem {inan} S. C. \.{I}nan, Phys. Rev. D {\bf 81}, 115002 (2010).
\bibitem {kepka} E. Chapon, C. Royon and O. Kepka, Phys. Rev. D {\bf 81}, 074003 (2010).
\bibitem {bil} S. Ata\u{g} and A. Billur, JHEP {\bf 11} 060 (2010).
\bibitem {bil2} \.{I}. \c{S}ahin and A. A. Billur, Phys. Rev. D {\bf 83}, 035011 (2011).
\bibitem {kok} \.{I}. \c{S}ahin and M. K\"{o}ksal, JHEP {\bf 11}, 100 (2011).
\bibitem {inan2} S. C. \.{I}nan and A. A. Billur, Phys. Rev. D {\bf 84}, 095002 (2011).
\bibitem {gru} R. S. Gupta, Phys. Rev. D {\bf 85}, 014006 (2012).
\bibitem {inanc2} \.{I}. \c{S}ahin, Phys. Rev. D {\bf 85}, 033002 (2012).
\bibitem{ban} B. \c{S}ahin and A. A. Billur,  Phys.Rev. D {\bf 85} 074026 (2012).
\bibitem {epl} L. N. Epele {\it et al.}, Eur. Phys. J. Plus {\bf127}, 60 (2012).
\bibitem {inanc3} \.{I}. \c{S}ahin and B. \c{S}ahin, Phys. Rev. D {\bf 86}, 115001 (2012).
\bibitem {bil4} A. A. Billur, Europhys. Lett. {\bf101}, 21001 (2013).
\bibitem {inanc4} \.{I}. \c{S}ahin {\it et al.}, Phys.Rev. D {\bf 88} 095016 (2013).
\bibitem{hao1} H. Sun, C. X. Yue, Eur. Phys. J. C 74, 2823 (2014).
\bibitem{hao2} H. Sun, Nucl. Phys. B 886, 691 (2014) arXiv:1402.1817 [hep-ph].
\bibitem{sen} A. Senol, A. T. Tasci, I. T. Cakir and O. Cakir, arXiv:1405.6050 [hep-ph].
\bibitem{kok2} M. K\"{o}ksal and S. C. \.{I}nan, Adv. High Energy Phys. 2014, 315826 (2014)
\bibitem{ha1} H. Sun, arXiv:1407.5356.
\bibitem{ha2} H. Sun, Y. J. Zhou and H. S. Hou, arXiv:1408.1218.
\bibitem{tas} M. Tasevsky, arXiv:1407.8332 [hep-ph].
\bibitem{ins} \.{I}. \c{S}ahin {\it et al.}, arXiv:1409.1796.


\bibitem{ep} M.S.Chen, T.P.Cheng, I.J.Muzinich and H.Terazawa, Phys.Rev.D, {\bf 7}, 3485 (1973).
\bibitem{budnev} V. Budnev, I. Ginzburg, G. Meledin and V.
Serbo, Phys. Rep. {\bf 15}, 181 (1975).
\bibitem{baur} G. Baur {\it et al.}, Phys. Rep. {\bf 364}, 359 (2002).
\bibitem{koh} V. A. Khoze, A. D. Martin and M. G. Ryskin, Eur. Phys. J. C 23, 311 (2002).
\bibitem{atsb} G. Aad {\it et al.} ATLAS Collaboration , ATLAS-CONF-2011-089 (2011).
\bibitem{cmsb} S. Chatrchyan {\it et al.} CMS Collaboration, J. Inst. 8 P04013 (2013).
\bibitem{fav} J. de Favereau de Jeneret, {\it et al.}, arXiv:0908.2020 [hep-ph].
\bibitem{pie} T. Pierzchala and K. Piotrzkowski, Nucl. Phys. Proc. Suppl. 179-180, 257 (2008)
arXiv:0807.1121 [hep-ph].
\bibitem{par} K.A. Olive et al. (Particle Data Group), Chin. Phys. C, 38, 090001 (2014).
\bibitem{cmss} CMS Collaboration, CMS-PAS-FTR-13-016.
\bibitem{ats} ATLAS Collaboration, ATL-PAHYS-PUB-2012-001.
\bibitem{sa1} J. A. Aguilar-Saavedra, Nucl. Phys. B {\bf812} (2009).
\bibitem{sa2} J. A. Aguilar-Saavedra, Nucl. Phys. B {\bf837} (2010).
\bibitem{ma} G. Mahlon, arXiv:0011349
\end{thebibliography}
\end{document}